\documentclass{ws-jai}

\usepackage[flushleft]{threeparttable}
\usepackage{float} 
\usepackage[dvipsnames]{xcolor}
\usepackage{caption}
\usepackage{physics}
\usepackage{amsfonts}
\usepackage{soul}

\usepackage{aas_macros}    

\usepackage[bookmarks,bookmarksnumbered,%
            urlcolor=blue,citecolor=blue,linkcolor=blue,%
            colorlinks=true,hyperfigures=true,hyperindex=true
           ]{hyperref}
           
\usepackage[normalem]{ulem}

\begin{document}

\catchline{}{}{}{}{} 

\markboth{Kwak et al.}{The Effects of the Local Environment on a Compact Radio Interferometer I}

\title{The Effects of the Local Environment on a Compact Radio Interferometer I: \\ Cross-coupling in the Tianlai Dish Pathfinder Array
}

\author{Juhun Kwak$^{1,7}$, John Podczerwinski$^{1}$, Peter Timbie$^{1}$, R\'eza Ansari$^{2}$, John Marriner$^{3}$, Albert Stebbins$^{3}$,\\ Fengquan Wu$^{4}$, Haotian Cao$^{1}$, Xuelei Chen$^{4,5,6}$, Kai He$^{4}$, Jixia Li$^{4,5}$, Shijie Sun$^{4,5}$, Jiacong Zhu$^{4}$}

\address{
$^{1}$Department of Physics, University of Wisconsin -- Madison, Madison, Wisconsin 53706, USA\\
$^{2}$Universit\'e Paris-Saclay, CNRS/IN2P3, IJCLab, 91405 Orsay, France\\
$^{3}$Fermi National Accelerator Laboratory, P.O. Box 500, Batavia IL 60510, USA\\
$^{4}$National Astronomical Observatory, Chinese Academy of Sciences, 20A Datun Road, Beijing 100101, China\\
$^{5}$University of Chinese Academy of Sciences, Beijing 100049, China\\
$^{6}$Center of High Energy Physics, Peking University, Beijing 100871, China}

\maketitle

\corres{$^{7}$dk.lightchaser@gmail.com}

\begin{history}
\received{(to be inserted by publisher)};
\revised{(to be inserted by publisher)};
\accepted{(to be inserted by publisher)};
\end{history}

\begin{abstract}
\textbf{Abstract:}
The visibilities measured by radio astronomical interferometers include non-astronomical correlated signals that arise from the local environment of the array. These correlated signals are especially important in compact arrays such as those under development for 21\,cm intensity mapping. The amplitudes of the contaminated visibilities can exceed the expected 21\,cm signal and represent a significant systematic effect. We study the receiver noise radiated by antennas in compact arrays and develop a model for how it couples to other antennas.  We apply the model to the Tianlai Dish Pathfinder Array (TDPA), a compact array of 16, 6-m dish antennas.  The coupling model includes electromagnetic simulations, measurements with a network analyzer, and measurements of the noise of the receivers. We compare the model to drift-scan observations with the array and set requirements on the level of antenna cross-coupling for 21\,cm intensity mapping instruments. { We find that for the TDPA, cross-coupling would have to be reduced by TBD orders of magnitude in order to contribute negligibly to the visibilities.}
\end{abstract}


\section{Introduction}
21\,cm intensity mapping is a technique for measuring the large scale structure of the Universe using the redshifted 21\,cm line from neutral hydrogen gas (HI) \cite{Liu&Shaw2020, Morales&Wyithe2010}. It is an example of the general case of line intensity mapping \cite{Kovetz2019}, in which spectral lines from any species, such as CO and CII, are used to make three-dimensional, ``tomographic" maps of large cosmic volumes. 21\,cm intensity mapping is used to study the formation of the first objects during the Cosmic Dawn and the Epoch of Reionization (6 $\leq$ z $\leq$ 50) and for addressing other cosmological questions with observations in the post-reionization epoch (z $\leq$ 6), such as constraining inflation models \cite{Xu2016} and the equation of state of dark energy \cite{Xu2015}. In the latter epoch, the approach provides an attractive alternative to galaxy redshift surveys. It measure the collective emission from many haloes simultaneously, both bright and faint, rather than cataloging just the brightest objects. As a result, the required angular resolution is relaxed as individual galaxies do not need to be resolved. By observing with wide-band receivers one simultaneously obtains signals over a range of redshifts and can construct a tomographic map. The primary analysis tool for cosmological measurements is the three-dimensional power spectrum.  Intensity mapping is a natural means to compute this spectrum over a range of wavenumbers, $k$, in which the perturbations are in the linear regime. Of particular interest in the power spectrum are the baryon acoustic oscillation (BAO) features, which can be used as a cosmic ruler for studying the expansion rate of the Universe as a function of redshift.

21\,cm intensity mapping, however, is challenging for various reasons. The primary concern is that the HI signal is orders of magnitude weaker than other radio sources. The main contaminants are galactic and extragalactic foregrounds, which are difficult to remove accurately. However, a wide variety of techniques have been studied for approaching this problem \cite{ansari2012, ewall-wice2020, chen2022, marins2022}. While it is true that the astrophysical foregrounds dominate the expected HI signal, radio emission from the environment in the vicinity of the radio telescope can also dwarf the HI signal. One such source is thermal emission from the ground, and we will study it in a future paper. Another source, the focus of this paper, is thermal noise emitted by the telescope receivers. Cross-coupling, also referred to as mutual coupling, or crosstalk, of receiver noise between antennas produces non-zero visibilities. The study of cross-coupling is partially motivated by our observations with the Tianlai Dish Pathfinder Array (TDPA) \citep{Wu2021} in which we found that the mean visibility for any given baseline over 24 hours of observation is nonzero and fairly stable from night to night (Fig. 26 of that paper). The nightly mean visibilities have a magnitude of 10's of mK and we subtract them from each baseline.  However, these nightly means are not perfectly stable and their fluctuations ultimately prevent the receiver noise from integrating down over periods of several days for Fourier modes along the line of sight with small values of $k_{||}$ (Fig. 32 of that paper). We study below the contribution of the cross-coupled receiver noise. While Fig. 33 of \citet{Wu2021} briefly addresses this topic, that study was not nearly as detailed and systematic as the work presented here. 


The term ``cross-coupling" is often used to refer to a broader class of instrumental effects in the literature. In \citet{Kern2019, Kern2020}, the authors use cross-coupling to refer to the sky signal that is reflected by the array elements and picked up by other antennas in the Hydrogen Epoch of Reionization Array (HERA). They studied this effect in delay space and attempt to calibrate it with the Singular Value Decomposition (SVD) technique. \citet{Fagnoni2021} and \citet{Josaitis2022} extend the analysis to study how this cross-coupling affects the beam patterns in HERA. They use a semi-analytic approach with single antenna beam patterns and a model for the interaction between dishes. Kern studied the reflections of sky signals between the two antennas
in a single baseline while Josaitis and Fagnoni extended the model to
include reflections off of all antennas in the array (not just the two antennas
in a particular baseline). Fagnoni studied this using electromagnetic
simulations while Josaitis studied this with a semi-analytic model. \citet{Ung2020} investigate the effect of mutual coupling between antennas in the Murchison Widefield Array (MWA) and the Engineering Development Array (EDA) on the noise temperature of the receivers. The present paper also studies receiver noise, but uses cross-coupling to refer specifically to thermal noise from receivers that is radiated from the antennas and picked up by other antennas in the array. In this work, we simulate electromagnetically the strength of antenna coupling and compare it to observations and direct measurements with the TDPA to provide a model and a calibration strategy for the effect. 
We expect that our methods and findings are sufficiently general and efficient that they could be applied to any line intensity mapping experiment using radio interferometers with appropriate but simple modifications. A similar approach was used by \citet{Sun2022} for the Tianlai Cylinder Pathfinder Array; the present paper introduces a more detailed model of the receiver noise. 

The rest of the paper is organized as follows: in Sect. \ref{TDPA}, we summarize the characteristics of the TDPA. In Sect. \ref{s:model}, we describe our models for receiver noise and cross-coupling. In Sect. \ref{CST}, we introduce CST Studio Suite, an electromagnetic simulation software package that we use to calculate the cross-coupling, and describe how we perform the simulations. Sect. \ref{measurements} outlines the measurements we have made of the receivers and antennas of the TDPA. In Sect. \ref{result}, we analyze our results from the cross-coupling simulations in comparison with the data from direct measurements and observations. Finally, in Sect. \ref{discussion}, we assess the magnitude of the effect and suggest measures that can be taken by future experiments to mitigate cross-coupling 
and conclude with Sect. \ref{conclusion}.

\section{Tianlai Dish Pathfinder Array}\label{TDPA}

The Tianlai program aims to make a 21\,cm intensity mapping survey of the northern sky \citep{Chen2012}. At present, the Tianlai program is in its Pathfinder stage to test the technology for making 21\,cm intensity mapping observations with an interferometer array. The Pathfinder consists of two arrays, one with 16 dish antennas, and the other with cylinder reflectors antennas (for the cylinder array, see \citet{Li2020a, Li2020b}), located next to each other at a radio quiet site (44$^\circ$9'N, 91$^\circ$48'E) in Hongliuxia, Balikun County, Xinjiang Autonomous Region, in northwest China. This paper presents the data from the dish array operation at 700 - 800\,MHz, corresponding to $1.03\geq z\geq 0.78$. The array 
will soon be re-tuned for observations in the 1330 - 1430\,MHz band ($0.07\geq z\geq -0.01$) to facilitate cross correlation with low-z galaxy redshift surveys and other low-z HI surveys. We summarize below the design of the dish array for the present purposes. More details about the TDPA can be found in \citet{Wu2021}.

\begin{figure}[h]
    \centering
    \includegraphics[width=0.9\textwidth]{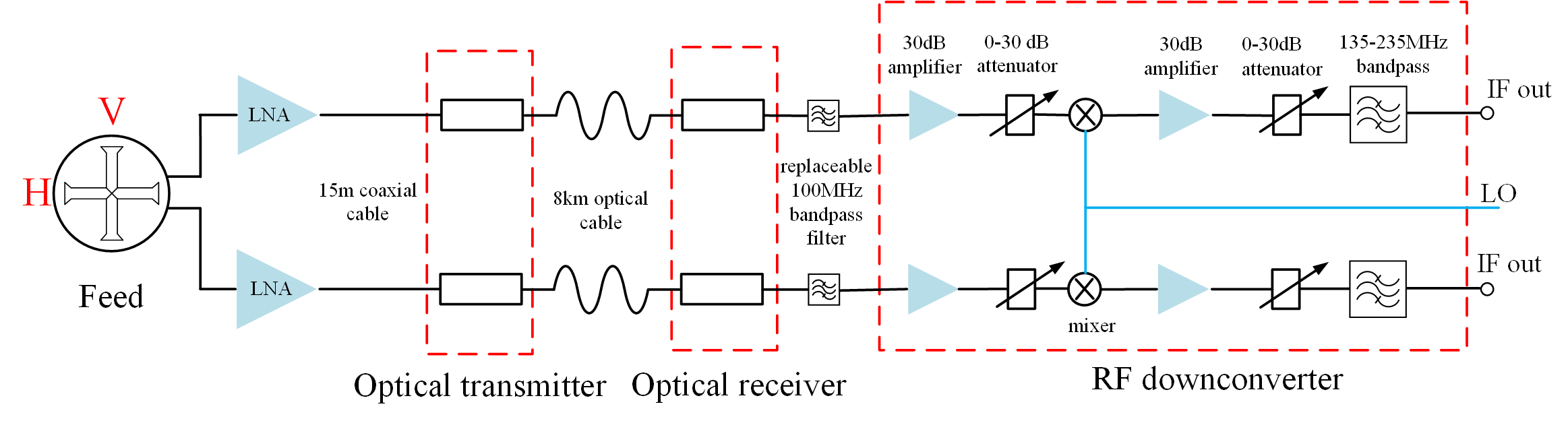}
    \caption{Schematic of the RF analog electronics.}
    \label{fig:AnalogSchematic}
\end{figure}

\begin{figure}[h]
    \centering
    \includegraphics[width=0.3\textwidth]{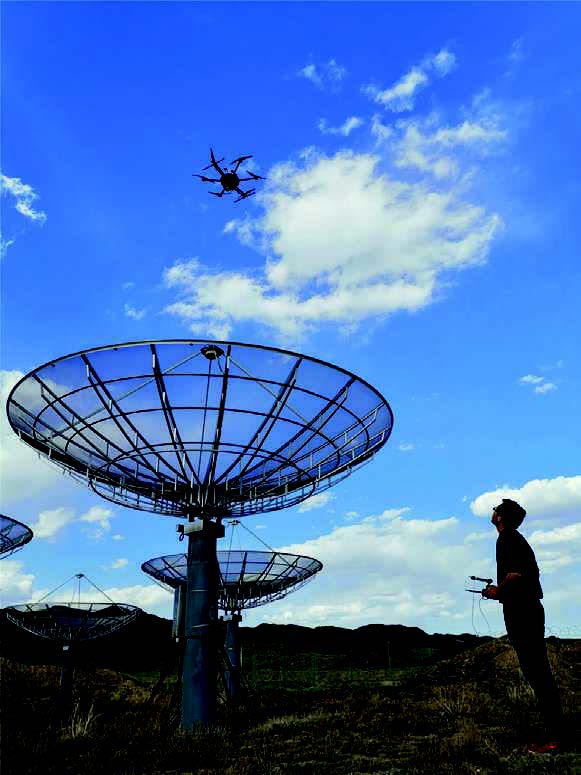}
    \includegraphics[width=0.3\textwidth]{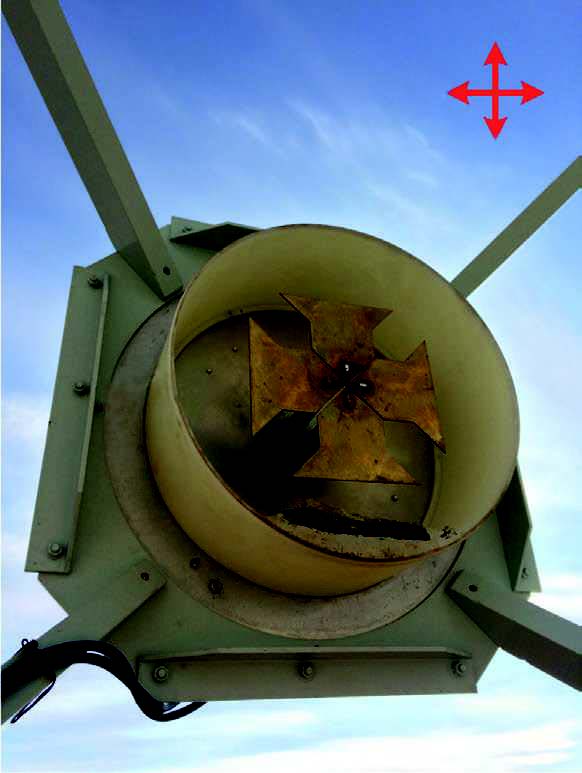}
    \caption{A dish antenna ({\bf left}) and feed antenna ({\bf right}) in the TDPA.}
    \label{fig:AntennaPhotos}
\end{figure}

The feed antennas, amplifiers, and reflectors are designed to operate from 400\,MHz to 1430\,MHz, corresponding to HI at the redshift of $2.55\geq z\geq -0.01$. The instrument operates with an RF bandwidth of 100\,MHz whose center can be tuned to any frequency in this range by adjusting the local oscillator frequency in the receivers and replacing the band pass filters. A schematic of the RF analog electronics appears in Fig. \ref{fig:AnalogSchematic}. The dish array consists of 16 on-axis dishes. Each has an aperture of 6\,m. The design parameters of the dishes are presented in Table \ref{tbl:antenna_param} and photographs of a dish antenna and feed antenna are in Fig. \ref{fig:AntennaPhotos}. The dishes are equipped with dual, linear-polarization receivers, and are mounted on Alt-Azimuth mounts. One polarization axis is oriented parallel to the altitude axis (horizontal, H, parallel to the ground) and the other is orthogonal to that axis (vertical, V) \cite{Zhang2021}. Motors are used to control the dishes electronically. The motors can steer the dishes to any direction in the sky above the horizon. The drivers are not specially designed for tracking celestial targets with high precision. Instead, in the normal observation mode, we point the dishes at a fixed direction and perform drift scan observations. The Alt-Azimuth drive provides flexibility during commissioning for testing and calibration. The dish array was fabricated by CASIC-23.

The dishes are currently arranged in a circular cluster. The array is roughly close-packed, with center-to-center spacings between neighboring dishes of approximately 8.8\,m. The spacing is chosen to allow the dishes to point down to elevation angles as low as 35$^\circ$ without ``shadowing" each other. One antenna is positioned at the center and the remaining 15 antennas are arranged in two concentric circles around it. It is well known that the baselines of circular array configurations are quite independent and have wide coverage of the $(u,v)$ plane. A comparison of the different configurations considered for the TDPA and the performance of the adopted configuration can be found in \cite{PAON4_Zhang_2016}. The Tianlai dishes are lightweight and the mounts are detachable, which enables the rearrangement of the antennas if needed. This paper describes observations with the array pointed either at the zenith or at the North Celestial Pole (NCP). The NCP region is a useful target because long integration times can be concentrated on a limited area of the sky. 

\begin{wstable}[h]
	\begin{minipage}{0.6\linewidth}
	    \caption{Main design parameters of a Tianlai dish antenna.}
        \centering
        \begin{tabular}{@{}ll@{}} \toprule
        Reflector diameter & 6\,m \\
        Antenna mount & Alt-Az pedestal \\
        f/D & 0.37 \\
        Feed illumination angle & 68$^{\circ}$ \\
        Surface roughness (design) & $\lambda/50$ at 21\,cm \\
        Altitude angle & 8$^\circ$ to 88.5$^\circ$ \\
        Azimuth angle & $\pm$360$^\circ$ \\
        Rotation speed of Az axis & 0.002$^\circ$ $\sim$ 1$^\circ$/s \\
        Rotation speed of Alt axis & 0.002$^\circ$ $\sim$ 0.5$^\circ$/s \\
        Acceleration & 1$^\circ$/s$^2$ \\
        Gain (design) & 29.4+20log(f/700\,MHz) dBi \\
        Total mass & 800 kg \\
        \botrule
        \end{tabular}
        \label{tbl:antenna_param}
	\end{minipage}\hspace{.5in}
	\begin{minipage}{0.4\linewidth}
		\centering
		\includegraphics[width=0.7\textwidth]{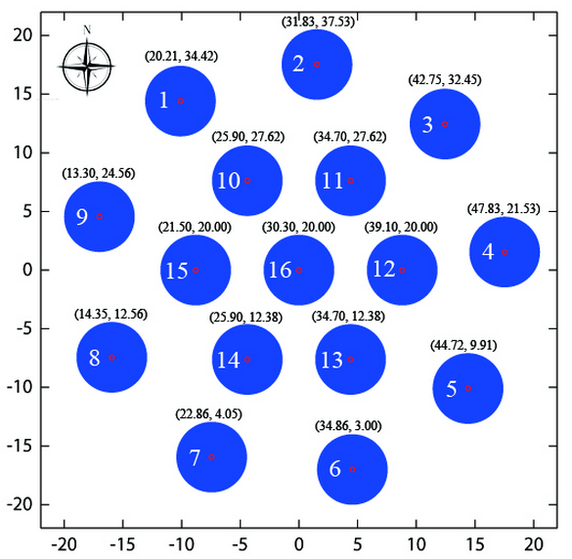}
		\captionof{figure}{TDPA in the configuration described above.}
		\label{fig:test}
	\end{minipage}
\end{wstable}

\section{Model}\label{s:model}
 
We apply the noise wave model of \citet{Meys1978} to estimate the noise emitted by the receivers toward the antennas. The dominant noise of each receiver is the LNA;  for simplicity, we consider only the LNA noise in the model.  For the TDPA case, noise from later stages is prevented from propagating backward to the LNA input by the LNA's high isolation (85 dB) in the reverse direction. This formalism is also used by the EDGES 21\,cm global spectrum instrument \cite{Monsalve2017}. 



\begin{figure}[t]
    \centering
    \includegraphics[width=0.9\textwidth]{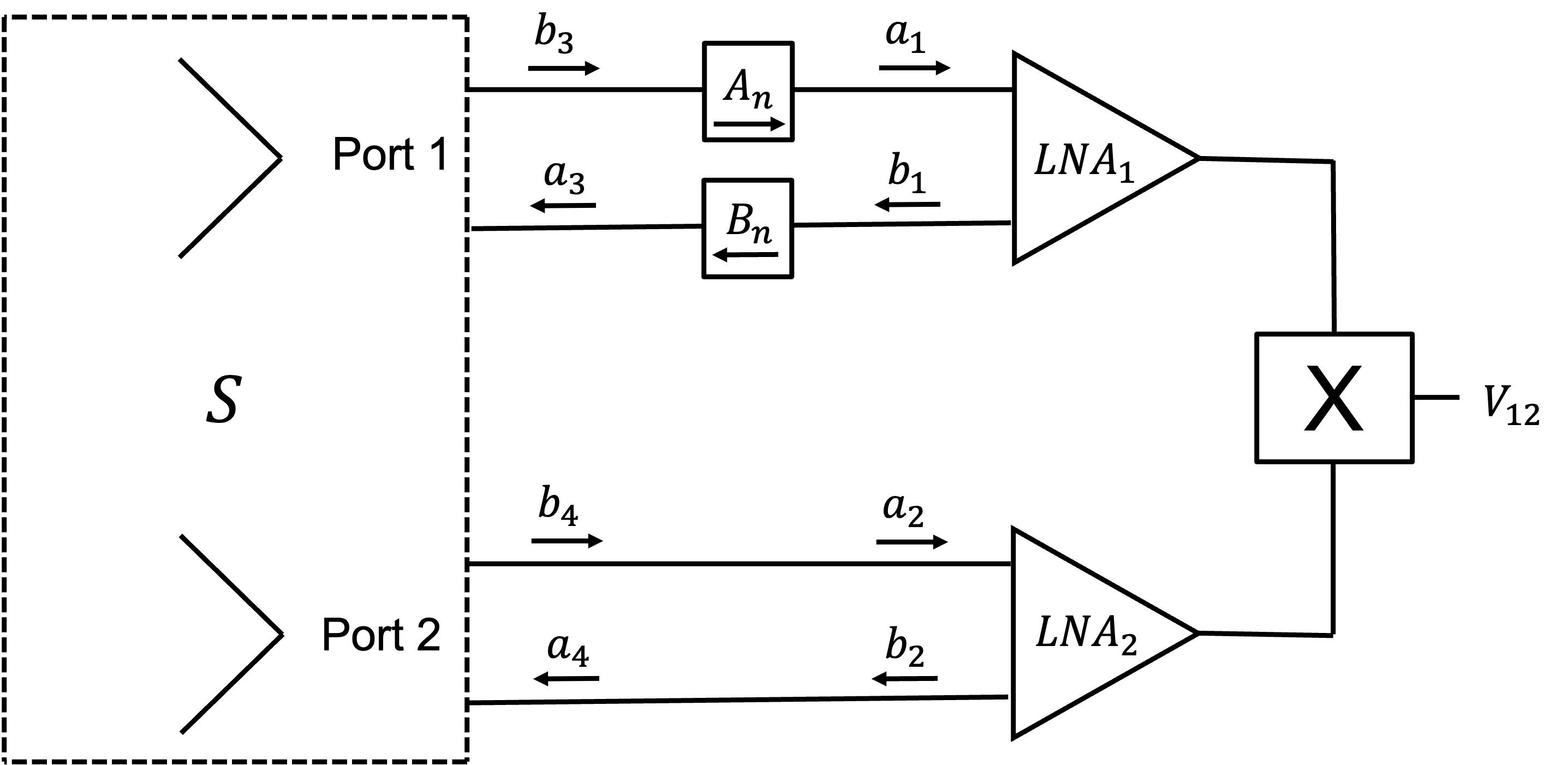}
    \caption{Amplifier Coupling Equivalent Circuit. Two low noise amplifiers, $\mathrm{LNA}_1$ and $\mathrm{LNA}_2$, with reflection coefficients $\Gamma_1$ and $\Gamma_2$, respectively, are connected by a network \textbf{S}, whose S-matrix is known and represents the coupling between a pair of antennas. Ports 1 and 2 are the output terminals of antennas 1 and 2, respectively.   The forward waves (the $a_i$'s) are shown for each device (LNAs and antenna ports) as are the backward waves (the $b_i$'s).  LNA$_1$ has two equivalent noise sources: a forward wave, $A_n$, and a backwards wave, $B_n$. 
    For clarity, only the noise generated by LNA$_1$ is shown. There are corresponding noise waves generated by LNA$_2$ that couple through the antennas into LNA$_1$.  
    The visibility formed by correlating the pair of output signals from the LNAs is $V_{12}$. }
    \label{fig:CouplingModel}
\end{figure}

A schematic of the model for receiver noise coupling between 2 antennas appears in Fig. \ref{fig:CouplingModel}. The receiver noise arises primarily from the low noise amplifiers (LNAs);  for simplicity, we consider only the LNA noise in the model. The two low noise amplifiers ($\mathrm{LNA}_1$ and $\mathrm{LNA}_2$) have a forward ($a$) and a reflected ($b$) wave at their respective inputs. These waves are related by complex reflection coefficients, $\Gamma_1$ and $\Gamma_2$, for the two LNAs, respectively.  The amplifiers are connected to the antennas, which form a network \textbf{S} described by an S-matrix.  The network \textbf{S} has 2 ports: each port has a forward wave ($a_3$ and $a_4$) and a reflected wave ($b_3$ and $b_4$) with the relation between forward and reflected waves given by the S-matrix equation:
\begin{equation}
\left( 
\begin{array}{c}
b_3\\
b_4 \\
\end{array} 
\right)
 = \left(
\begin{array}{cc}
S_{11} & S_{12} \\
S_{21} & S_{22} \\
\end{array} \right)
\left( 
\begin{array}{c}
a_3 \\
a_4 \\
\end{array} 
\right). \label{eqn:S}
\end{equation}
$\mathrm{LNA}_1$ contains noise sources, which are represented by an equivalent forward wave ($A_n$) and backward wave ($B_n$) at the $\mathrm{LNA}_1$ input. The two equivalent noise sources are the most general model of a linear 2-port amplifier: any combination of noise sources inside the amplifier can always be reduced to two equivalent noise wave sources. For simplicity, we consider here only the coupling of noise generated by $\mathrm{LNA}_1$ that passes through antenna 1 and into antenna 2 and $\mathrm{LNA}_2$.  $\mathrm{LNA}_2$ generates analogous noise waves that couple into antenna 1 and then $\mathrm{LNA}_1$; we account for the effect of $\mathrm{LNA}_2$ below. 

The two noise waves simply add to the forward and reflected waves. The reflected wave at LNA$_1$ is related to the forward wave by the reflection coefficient $\Gamma_1$. The reflected wave at the input of \textbf{S} is given by the S-matrix parameters and depends on the forward wave at the opposite port ($a_4$). The equations describing the network are:  
\begin{align}
a_1 &= A_n + b_3 \\
b_1 & = \Gamma_1 a_1 \\
a_3 & = b_1 + B_n \\
b_3 & = S_{11} a_3 + S_{12} a_4 .
\end{align}
The term proportional to $a_4$ describes the noise that is coupled to LNA$_2$ and reflected back to LNA$_1$. We assume the case where $S_{12}$ is small and $S_{12}=S_{21}^*$, so that the term involving $a_4$ is second order in the coupling and can therefore be neglected. If we were to include that term, we would have to include all other devices that could provide similar reflections back to LNA$_1$. In the approximation $S_{12}=0$, the input circuit of LNA$_1$ is decoupled from the input of LNA$_2$, and the 4 equations can be solved with the result:

\begin{equation}
a_1 = \frac{A_n + S_{11} B_n}{1 - \Gamma_1 S_{11}}
\end{equation}
\begin{equation}
a_3 = \frac{\Gamma_1 A_n +  B_n}{1 - \Gamma_1 S_{11}}\\[5 pt]
 \end{equation} 
The forward wave at LNA$_1$ is the forward noise wave plus the reflected portion of the backward wave. The denominator is a resonant effect, which is small unless $\Gamma_1$ and $S_{11}$ are both close to 1. A similar expression holds for the forward wave $a_3$. We can now solve for the input circuit of LNA$_2$. The relevant equations are: 
\begin{align}
a_2 &= b_4 \\
b_2 & = \Gamma_2 a_2 \\
a_4 & = b_2 \\
b_4 & = S_{21} a_3 + S_{22} a_4
\end{align}
and the solution for $a_2$ is
\begin{align}
a_2 & = \frac{S_{21} a_3}{1-\Gamma_2 S_{22}} \\
& = \bigg[ \frac{S_{21}}{1-\Gamma_2 S_{22}} \bigg] \bigg[\frac{B_n + \Gamma_1 A_n}{1-\Gamma_1 S_{11}} \bigg].
\end{align}
The contribution to the visibility $V_{12}$ from the noise of LNA$_1$ is (neglecting the gains of the LNAs):
\begin{equation}
V_{12} = \langle a_2 a_1^* \rangle = \bigg[\frac{S_{21}}{1 - \Gamma_2 S_{22}} \bigg] \bigg \lvert \frac{1}{1-\Gamma_1 S_{11}} \bigg \rvert ^2 
 \bigg[ \Gamma_1 \langle \abs{A_n}^2 \rangle + \Gamma_1 S_{11}^*\langle A_n B_n^*\rangle + \langle A_n^* B_n\rangle  + S_{11}^* \langle \abs{B_n}^2 \rangle \bigg].
 \label{Eq:visibility}
\end{equation}


\noindent The terms in this equation are defined (after equation 5 of \citealt{Meys1978}) in terms of the noise temperatures of the LNA:
\begin{equation}
    \langle \abs{B_n}^2 \rangle = kT_b\Delta f,
    \label{Eq:B}
\end{equation}
\begin{equation}
    \langle\abs{A_n}^2\rangle = kT_a\Delta f,
    \label{Eq:A}
\end{equation}
\begin{equation}
    \langle A_n^* B_n \rangle =  kT_c e^{i \phi_c}\Delta f,
    \label{Eq:AB}
\end{equation} 
\begin{equation}
    S_{11} = \abs{S_{11}}e^{i\phi_{s}},
    \label{Eq:S}
\end{equation}
where $k$ is Boltzmann's constant and $\Delta f$ is the RF bandwidth. 
These three noise temperatures, $T_a$, $T_b$, and $T_c$, the phase factor $\phi_c$, and the reflection coefficient $\Gamma_1$ are properties of the LNA. They were measured for one of the TDPA LNAs following the procedure described below, in Sec. \ref{measurements}. We find that $\Gamma_1 \sim S_{11}$ are small and in the following we only work to the first order of these quantities. 
The quantities $S_{21}$ and $S_{11}$ are both measured and simulated, as described below. With these definitions and approximations,
\begin{equation}
    \left\langle \abs{V_{12}}^2\right\rangle \approx  \abs{S_{21}}^2 (k\Delta f)^2 T_c\,\mathfrak{Re}\left[T_ce^{2i\phi_c}+2\Gamma_1T_ae^{i\phi_c}+2\abs{S_{11}}T_be^{i(\phi_c-\phi_s)}\right].
     \label{Eq:Visibility_temp_units_1}
\end{equation}
Note that we have only considered the visibility produced by the cross-coupled receiver noise from LNA$_1$ to LNA$_2$. The total contribution to the visibility is doubled, assuming that the crosstalk of receiver noise from LNA$_2$ to LNA$_1$ is the same as that from LNA$_1$ to LNA$_2$. Therefore, the total contribution to the visibility from the cross-coupled receiver noise from is, in temperature units,
\begin{equation}
    V_{12,T} = 2\sqrt{\left\langle\abs{V_{12}}^2\right\rangle/(k\Delta f)^2}.
    \label{Eq:Visibility_temp_units_2}
\end{equation}
Note also that we are working only to first order in $S$, so the noise from one antenna that couples to other pairs of antennas (not shown in Fig. \ref{fig:CouplingModel}) are neglected. However, this effect still has finite contributions to visibilities formed by each of the pairs.

\section{Electromagnetic Simulations}\label{CST}

\subsection{Cross-coupling in CST}\label{CSTXcoupling}


To simulate the effect of cross-coupling,
we treat the TDPA as a network, \textbf{S}, similar to that defined in Section \ref{s:model}, but now including 32 ports (one for each polarization of each antenna) instead of two. The ports are defined to be at the interface between the antennas and the LNAs. We simulate the outgoing voltage wave at a port of interest to the voltage incident on one of the other ports in the array using a scattering matrix:
\begin{equation}
    \begin{bmatrix}
        b_1 \\ b_2 \\ \vdots \\ b_{32}
    \end{bmatrix}
    =
    \begin{bmatrix}
        S_{11} & S_{12} & \cdots & S_{1~32} \\
        S_{21} & & & \vdots \\
        \vdots & & & \vdots \\
        S_{32~1} & \cdots & \cdots & S_{32~32}
    \end{bmatrix}
    \begin{bmatrix}
        a_1 \\ a_2 \\ \vdots \\ a_{32}
    \end{bmatrix}
\end{equation}
which allows for the calculation of a specific element of the scattering matrix as
\begin{equation}
    S_{ij} = \frac{b_i}{a_j}\Big|_{a_k=0 \ \mbox{for} \ k \not= j}.
\end{equation}

\begin{figure}[h]
    \centering
    \includegraphics[width=0.3\textwidth]{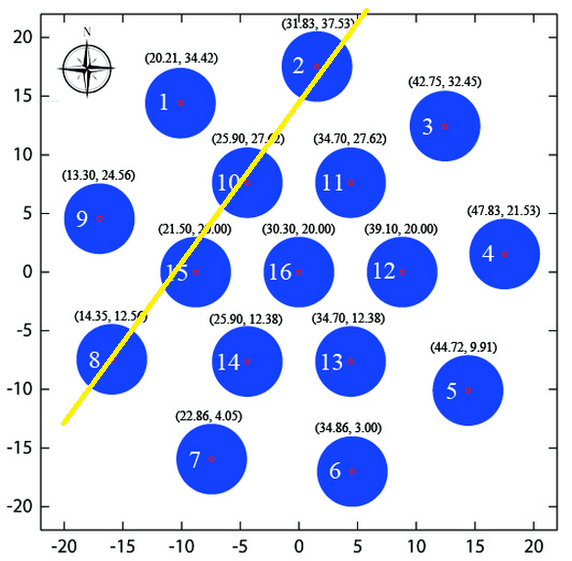}
    \caption{Baselines whose cross-coupling is analyzed in detail in this paper are 2V-10V, 2V-15V and 2V-8V. V stands for the ``vertical" polarization, oriented along the North-South direction. The coordinates of the dish centers are given on the figure in units of meters. 
    }
    \label{fig:selected_baselines}
\end{figure}
\begin{figure}[h]
    \centering
    \includegraphics[width=0.7\textwidth]{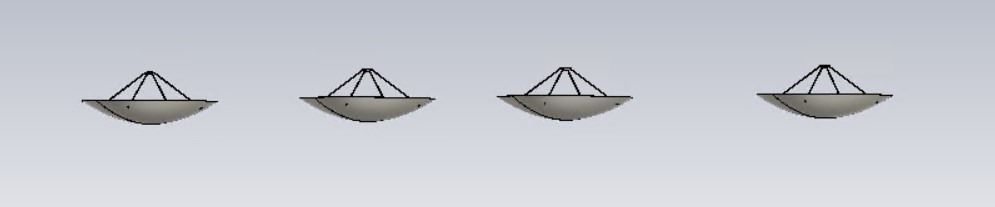}
    \includegraphics[width=0.7\textwidth]{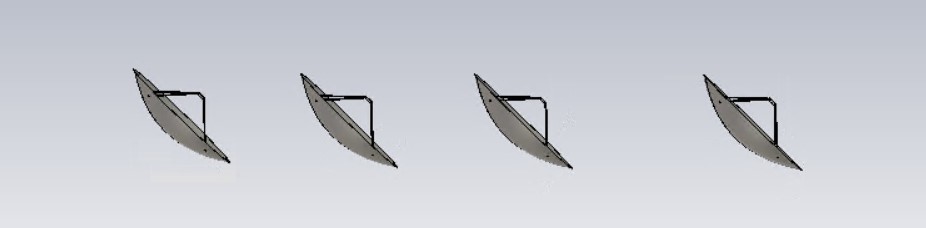}
    \caption{ Renderings of a subset of the 16 dishes simulated using CST, with dishes pointing at the zenith (\textbf{top}) and the NCP (\textbf{bottom}). $S_{21}$ between pairs of dishes is lower in the NCP case because the dishes block the line-of-site path between feed antennas. The dish numbers are 2, 10, 15, and 8, from right to left. }
    \label{fig:CSTrenderings}
\end{figure}

To compute the elements of \textbf{S} we use CST Studio Suite. CST is a commercial electromagnetic simulation software package 
that is particularly useful for problems that include antennas. 
In our simulations, we point all 16 dishes of the array toward the North Celestial Pole (NCP) to compare to the existing observational data from the TDPA. The off-zenith angle is 45.1$^\circ$ when the array points at the NCP.  Renderings of the geometry of the CST simulations in the zenith-pointing case and the NCP-pointing case are included Fig. \ref{fig:CSTrenderings} to visualize the light paths. The simulation of each antenna includes both the reflector and the feed antenna, both of which are described in \citet{Zhang2021}. All materials that make up the array are set to be perfect electric conductor (PEC) for simplicity; the ground is neglected. The cross-coupling in the baselines of 2V-10V, 2V-15V and 2V-8V, as highlighted in Fig. \ref{fig:selected_baselines}, is presented in detail in this paper. There are different solvers available in CST; we use the integral equation (IES) solver for simulations of the TDPA presented in this work. A description of selected solvers and the reasoning behind choosing them is explained in detail in the appendix.

\begin{figure}[h]
    \centering
    \includegraphics[width=0.85\textwidth]{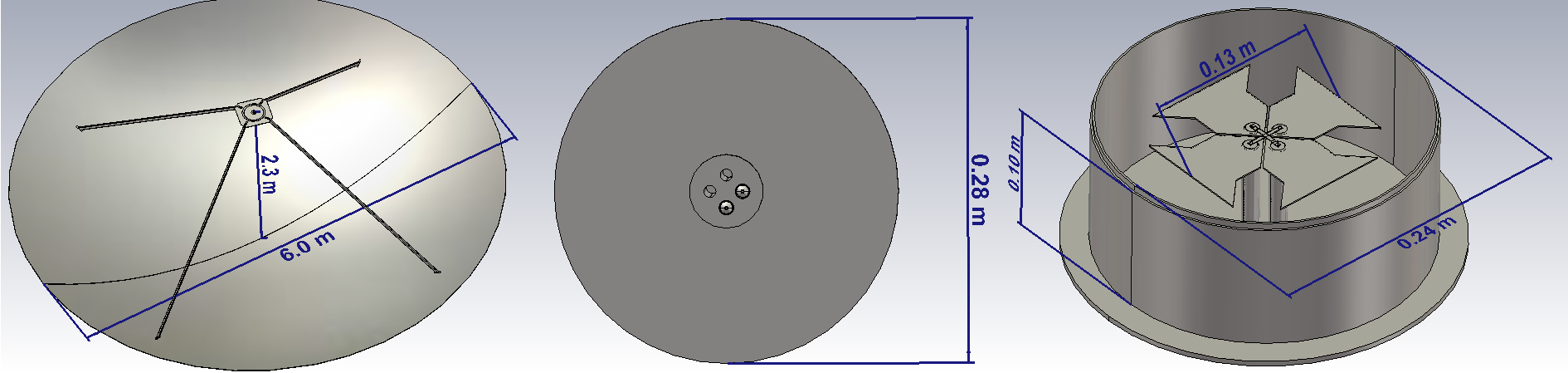}
    \caption{ Detailed rendering of the antenna design used in the simulations. ({\bf Left}) shows the feed when placed over the dish. ({\bf Center}) shows the backplate and ports of the feed antenna. ({\bf Right}) shows the feed dipoles. All elements shown in these renderings are assumed to be made of PEC.}
    \label{fig:enter-label}
\end{figure}

\subsection{Computation Cost}
Simulations of cross-coupling with CST can be computationally expensive and time-consuming. The RAM requirement and simulation time are typically the limiting factors. These depend particularly on 
the size of the array, which affects the number of mesh cells, which scales inversely as frequency.  Other factors, such as material types, are important as well. The size of our array is a circle of radius $\sim 20$\,m with 16 dishes, each of diameter $6$\,m.  For the mesh setting,  we use 15 cells per wavelength 
on surfaces and 5 in free space, which results in roughly five million cells for the entire array. With the medium accuracy setting, $10^{-3}$, which is the default for IES, this requires $\sim 300$\,GB of RAM. Because this is beyond the capacity of our lab computers we collaborate with the Center for High Throughput Computing (CHTC) at the University of Wisconsin - Madison to perform the simulations. The advantage of using computing clusters like CHTC is that, in addition to gaining access to computers with higher RAM, one can run many jobs simultaneously, as licenses allow. In our case, we break down a simulation with 101 frequency samples to 101 simulations with a single frequency sample and run $10 \sim 15$ of them in parallel. The parallelization has helped us to reduce the total simulation time required to less than a quarter of what would have taken if we used a single computer.

\subsection{Reliability}\label{SanityChecks}
{ To obtain the highest accuracy of the simulations, we experiment with the accuracy settings and the mesh settings. Unfortunately, we have not succeeded in running IES simulations at higher accuracy setting than medium due to the limitation on the computing resources. Even if there was a success, however, the computation time for the simulations involving the entire array is expected to be prohibitively long. The number of mesh cells that we use (15 cells per wavelength on the geometry and 3 cells on the open space) is found to be optimal in a test performed with two-dish zenith-pointing configuration.}

\begin{figure}[h]
    \centering
    \includegraphics[width=0.95\textwidth]{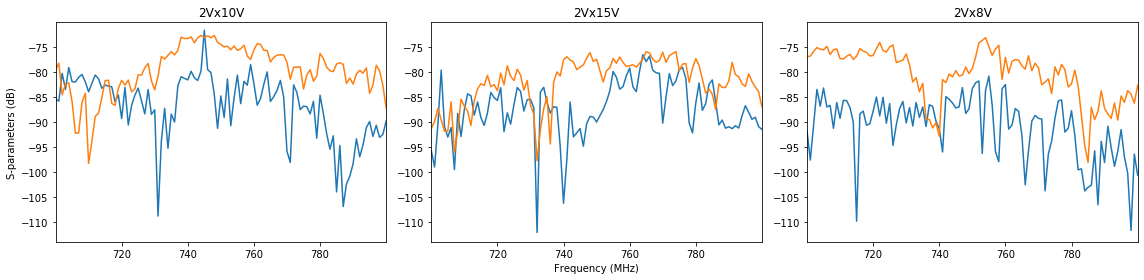}
    \caption{Simulated $S_{21}$ (blue) and $S_{12}$ (orange) for the corresponding baselines.}
    \label{fig:NCP-FA_symmetries}
\end{figure}

\begin{figure}[h]
    \centering
    \includegraphics[width=0.5\textwidth]{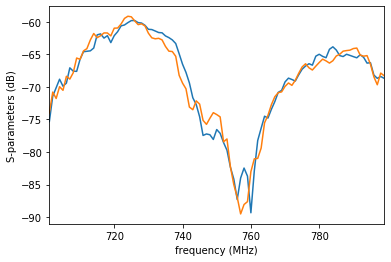}
    \caption{Simulated $S_{21}$ (blue) and $S_{12}$ (orange) with a pair of dishes pointing at zenith at 10\,m separation.}
    \label{fig:symmetry_pair_zenith}
\end{figure}

There are a few sanity checks that one can perform on the simulation result. One of the tests is the check of $S_{21}$ and $S_{12}$ symmetry. By symmetry, the response to port 2 to a signal from port 1 should be the same as the response of port 1 to the identical signal from port 2. One can find justification for this symmetry in chapters 11 and 12 of \citet{ramo}. However, the expected s-parameter symmetry is not seen in Fig. \ref{fig:NCP-FA_symmetries},
where all 16 dishes of the array are included and are pointing at the NCP. We therefore studied the symmetry of $S_{21}$ and $S_{12}$ in a simpler case, namely when a pair of dishes are pointing at zenith. Fig. \ref{fig:symmetry_pair_zenith} shows the result from this simulation. Numerically, the difference between $S_{21}$ and $S_{12}$ is $\sim 0.37 \%$ or $\sim 0.14$\,dB on average, showing the expected result.

\begin{figure}[h]
    \centering
    \includegraphics[width=0.95\textwidth]{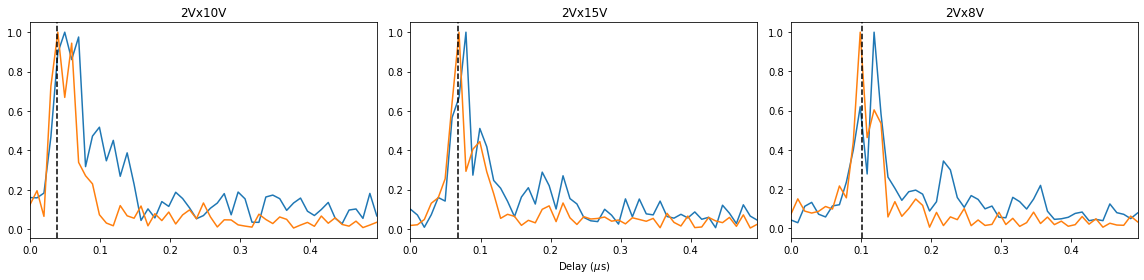}
    \caption{Delay spectra taken from simulated $S_{21}$ (blue) and $S_{12}$ (orange) of the corresponding baseline with all 16 dishes pointing at the NCP. The curves are normalized so the peak of each curve has magnitude of one. In reality, the orange curves have peaks at negative delay, consistent with the forward/backward traveling wave formalism. Here, they are plotted on the same side of the vertical axis as the blue curves for comparison. {The black dashed line marks the delays corresponding to the light travel time between dishes in the baseline, which matches quite well with the first dominant peak. The second dominant peak seems to have originated from the reflection between the feeds and the reflectors of the same antennas.} }
    \label{fig:DelaySanityCheck}
\end{figure}

Another test that one can perform is the delay spectrum analysis. One can gain confidence by checking whether the delay spectra formed from the simulated cross-coupling as a function of frequency show the physically sensible result, that is, having obvious peaks at the delays corresponding to the light travel time in each baseline. Fig. \ref{fig:DelaySanityCheck}, which presents the delay spectra taken from the cross-coupling simulation with the full array pointing at the NCP, shows that our results are physically sensible. The black dashed lines show where we expect the most prominent peaks to be, corresponding to the light travel time between the dishes. 
We think that the second peaks (about 20\,ns apart from the first peak, which is roughly the ratio of the dish diameter to speed of light) appear because of waves that reflect from the edge of a dish before coupling into a feed antenna. Later peaks are from waves that undergo additional reflections. 
Fig. \ref{fig:DelaySanityCheck} also shows that despite the broken $S_{21}$ and $S_{12}$ symmetry, the results make sense physically.

\begin{figure}[h]
    \centering
    \includegraphics[width=0.4\textwidth]{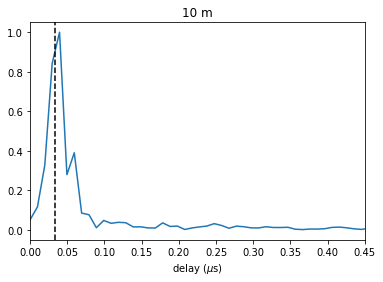}
    \hspace{0.02\textwidth}
    \includegraphics[width=0.4\textwidth]{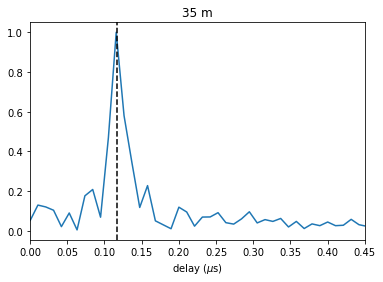}
    \caption{ Equivalent of Fig.\ref{fig:DelaySanityCheck} for the zenith pointing pair at the separation specified. We see predominant peaks at the delay corresponding to the separation between the pair, further building our confidence that the frequency structures in the cross-coupling are physical.}
    \label{fig:enter-label}
\end{figure}

{Repeating the same simulations does not seem to reproduce exactly the same values of cross-coupling at the same frequencies when the dishes are rotated away from the zenith.} We suspect that rotating the geometry compromises the accuracy of the CST simulation somewhat. {However, the delay spectra taken from different runs showed prominent peaks at the same delays, suggesting that the results of the simulations are physically sensible.}

\section{Measurements}\label{measurements}

\subsection{Receiver Noise Temperatures}

\begin{figure}[h]
    \centering
    \includegraphics[width=0.7\textwidth]{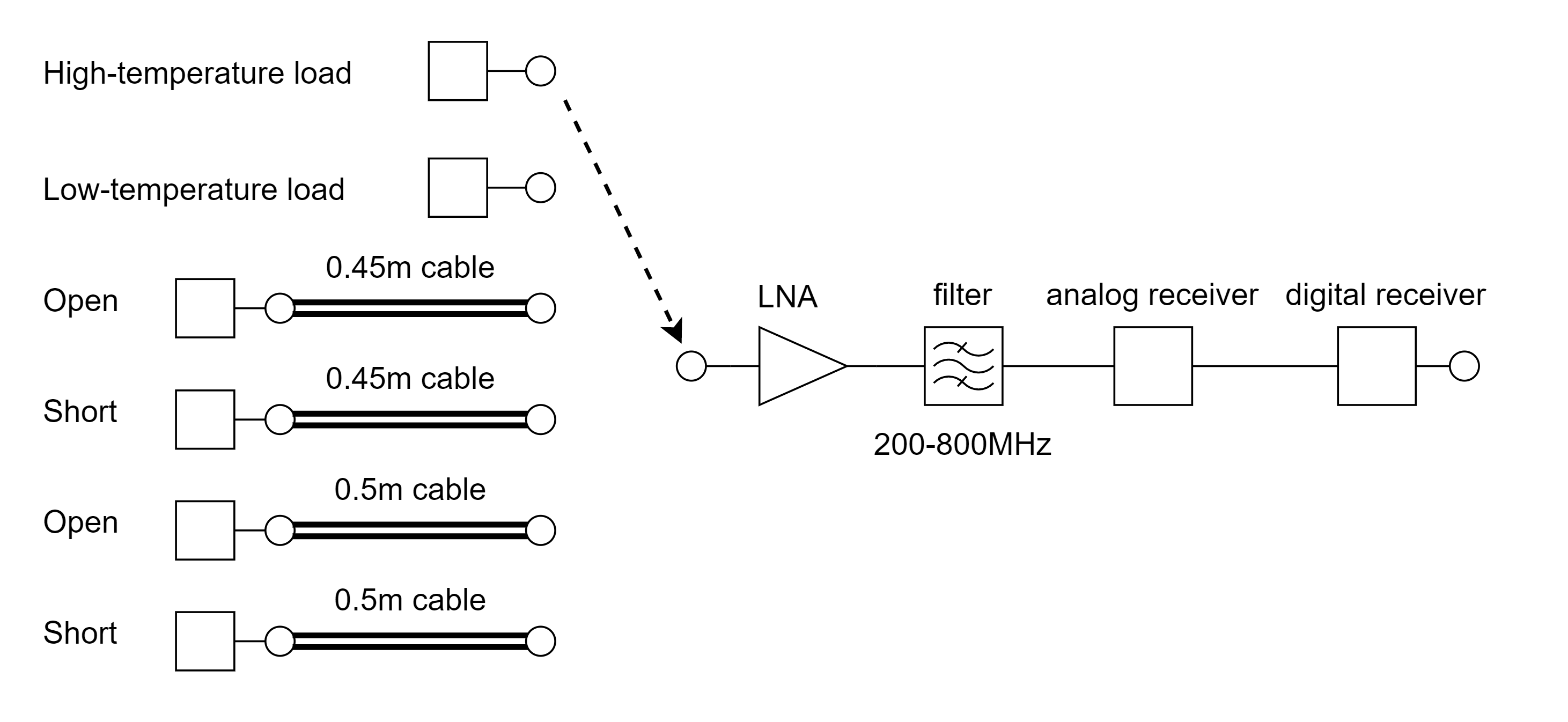}
    \caption{Experimental setup of noise temperature measurement.}
    \label{fig:LNA-noise-wave-measurement}
\end{figure}

The noise temperatures of the LNAs (Sect. \ref{s:model}) were measured for one of the LNAs in the same batch as those used in the TDPA receivers. The LNAs on the antennas are expected to be similar.  The measurement is made by observing the power at the output of the LNA as a function of RF frequency when the LNA input is connected to 6 different sources (loads with reflection coefficients $\Gamma_s = |\Gamma_s|e^{i\phi_s}$): open circuit and short circuit at the end of a cable of length $5\,$m; open circuit and short circuit at the end of a cable of length $0.45\,$m, and $50\,\Omega$ load maintained at a temperature $T_s$ which is either a high temperature ($\sim 390$K) or a low temperature (ambient)(see Fig. \ref{fig:LNA-noise-wave-measurement} for measurement setup). { The noise model of an amplifier with noise temperatures $T_a$, $T_b$, and $T_c$, and phase $\phi_c$  
is described in Sect. \ref{s:model}, and references \cite{meys_wave_1978, monsalve_calibration_2017}.  These parameters can be determined as a function of frequency by fitting to the noise power model $P$}: 


\begin{equation}
  P = [T_a + T_b|\Gamma_s|^2 + 2|\Gamma_s|[T_\text{cos} \cos{\phi_s} +T_\text{ sin}\sin{\phi_s}]+ T_s(1-|\Gamma_s|^2)] k\Delta f.
  \label{Eq:noisetest1}
\end{equation}
{ $P$ is the amount of power (``noise power") observed at the output of the amplifier in a bandwidth $\Delta f$ when the input is connected to a load with reflection coefficient $\Gamma_s$ and noise temperature $T_s$.  The power is referred to the amplifier input, so we neglect the amplifier gain.  (Effectively, the gain is measured and divided out.)} { $T_a$ is the portion of the LNA noise traveling forward into the LNA, $T_b$ is the portion of the LNA noise traveling backward to the antenna, and $T_c$ is the correlated portion of these forward and backward noise components: 
}
\begin{equation}
T_c \cos{\phi_c}\equiv T_\text{cos}
\label{Eq:noisetest2}
\end{equation}
and
\begin{equation}
  T_c \sin{\phi_c}  \equiv T_\text{sin}.
  \label{Eq:noisetest3}
\end{equation}

\begin{figure}[h]
    \centering
    \includegraphics[width=0.5\textwidth]{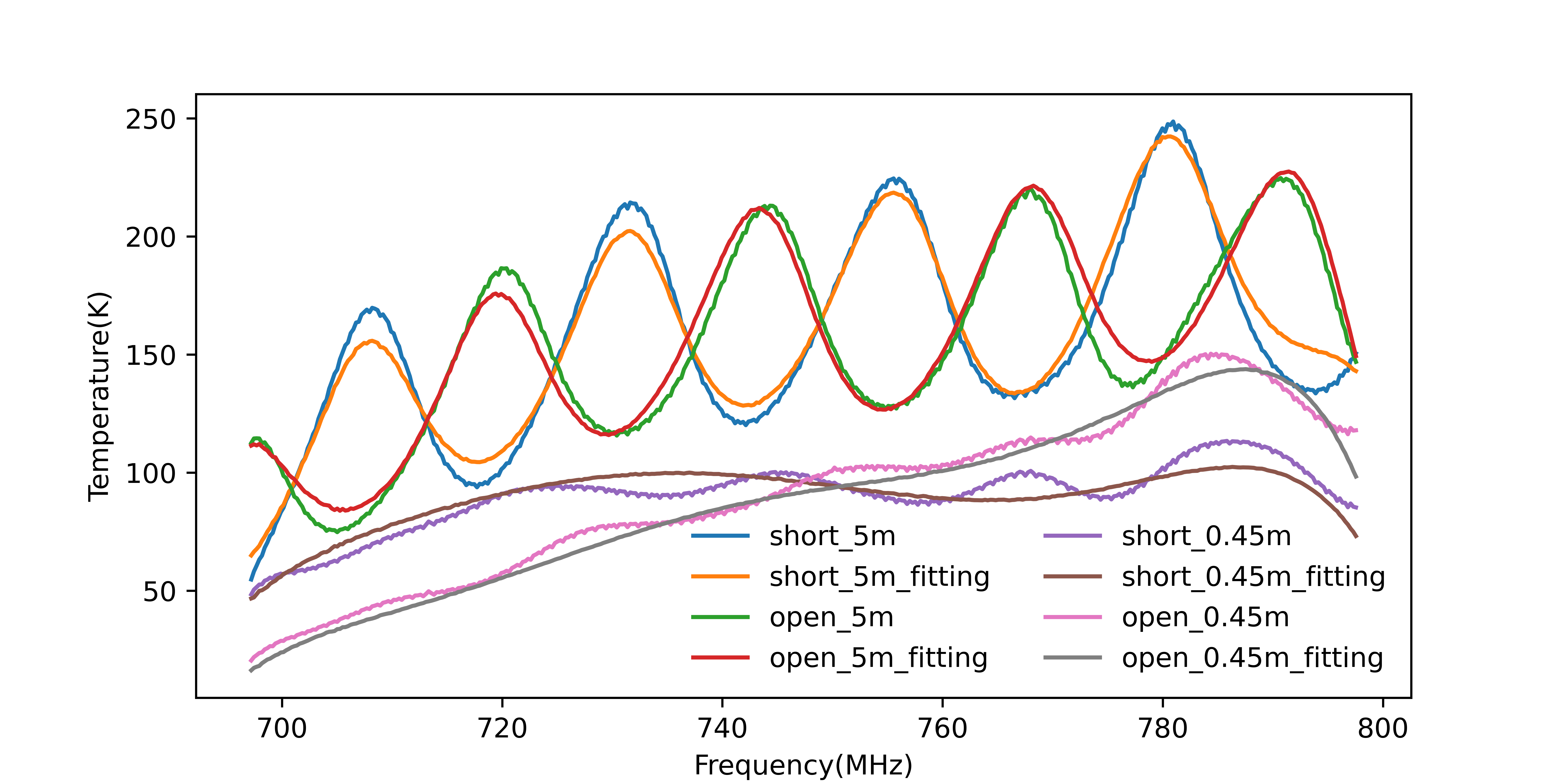}
    \caption{{Measured spectra and best-fitting results. The label ``short\_5m"  represents the measured spectrum with a 5-meter cable terminated in a short circuit in front of the LNA. 
    The label ``short\_5m\_fitting" represents the fitted model result for this spectrum. The labelling of the other curves are similar. 
    The temperature scale of the plot is calibrated from measurements of a $50\,\Omega$ load held at low-temperature and high-temperature.}  
    }
    \label{fig:NoiseWaveSpectra}
\end{figure}

{ The measured spectra from open and shorted cables are shown in Fig. \ref{fig:NoiseWaveSpectra}. The temperature scale on the ordinate has been calibrated in units of Kelvin by measurements of a low-temperature and high-temperature load.   The fitted results are also plotted for comparison.} 

\begin{figure}[h]
    \centering
    \includegraphics[width=0.5\textwidth]{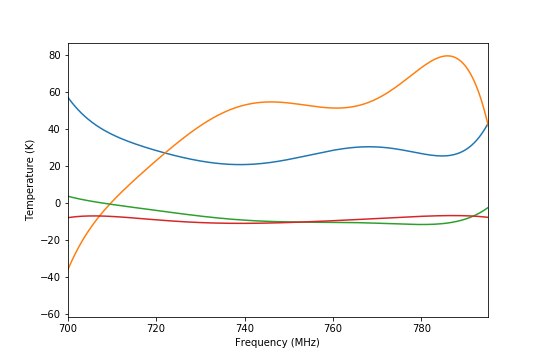}
    \caption{{Measured values for the noise temperatures that characterize the noise wave model of the LNA. They appear in Eq. \ref{Eq:noisetest1} - \ref{Eq:noisetest3}. The curves are:  $T_a$ (orange), $T_b$ (blue), $T_\text{cos}$ (green), and $T_\text{sin}$ (red).}}
    \label{fig:NoiseWaveParam}
\end{figure}

{
 The noise wave parameters $T_a$ (orange), $T_b$ (blue), $T_\text{cos}$ (green), and $T_\text{sin}$ (red) of the LNA are presented in Fig. \ref{fig:NoiseWaveParam}. 
 The frequency dependence of these parameters is assumed to be a 7th-order polynomial function in the fitting.
 Due to the bad frequency response of at the edge of the bandpass filter, the fitting may get some weird results at the edge of the band; for example, $T_a$ goes down to negative values below $710$\,MHz. But at the center part of the band, the results seem reasonable:  $T_a\approx 55$\,K, $T_b\approx 30$\,K, and $T_b$ is about one half of $T_a$.  The noise figure of the LNA we measured using a noise figure meter is about 0.7, which corresponds to $50$\,K. This noise figure measurement result is pretty close to $T_a$, as expected for a measurement with noise figure meter that is impedance-matched to an amplifier.
 For these measurements the integration time for each input source is about 20 minutes and the thermal noise is less than $0.01\,$K.}

\subsection{Cross-coupling}
\label{sec:cross-coupling}

\begin{figure}[h]
    \centering
    \includegraphics[width=1\textwidth]{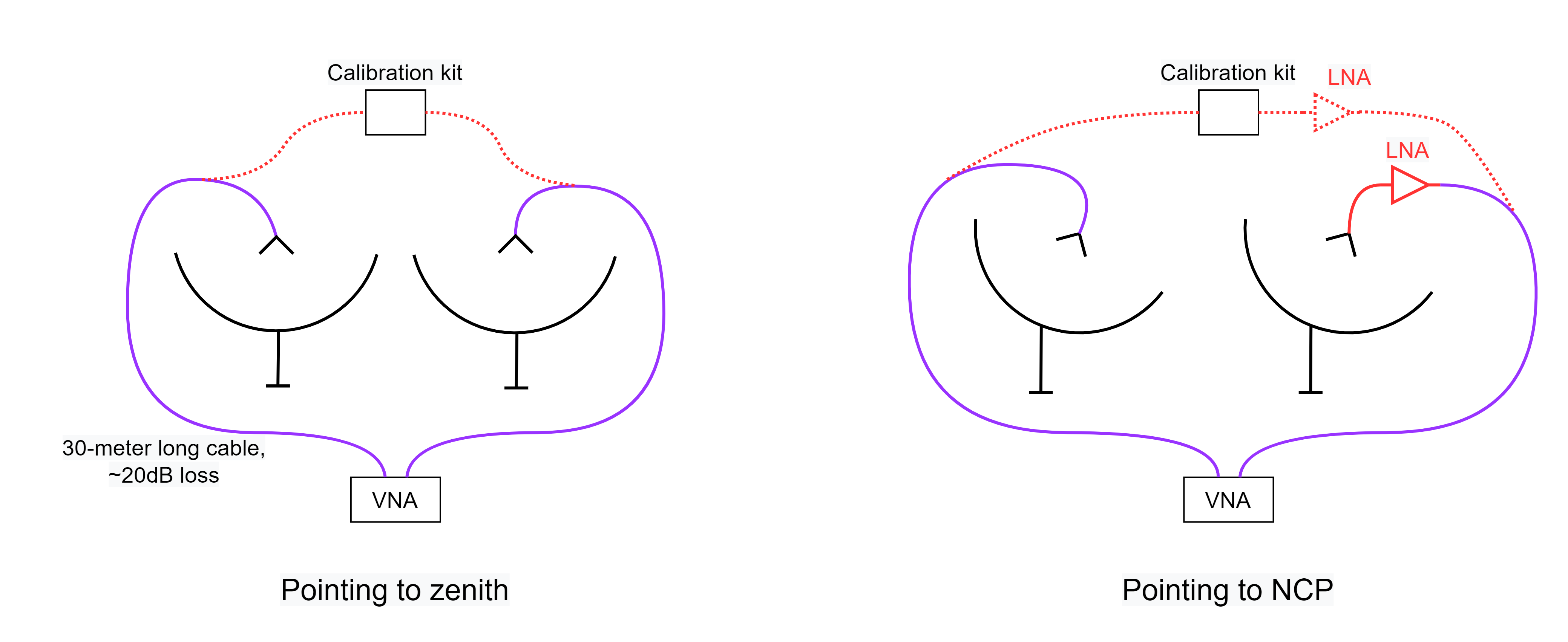}
    \caption{ Experimental setup of the cross-talk measurement of a pair of TDPA antennas when pointed at the zenith (left) and the NCP (right). The red dotted lines show the configuration of the calibration kit during calibration of the VNA for the zenith and NCP measurements.}
    \label{fig:VNA-crosstalk-measurement}
\end{figure}

Measurements of the $S_{21}$ and $S_{11}$ scattering parameters were made of baselines 2V-8V, 2V-10V, and 2V-15V of the TDPA using a vector network analyzer (VNA, Copper Mountain TR1300). 
{  The measurement setup can be found in Fig. \ref{fig:VNA-crosstalk-measurement}. } 
The VNA was connected to pairs of antennas through cables of length $30\,$m.  The loss in the cables was about $-20\,$dB. The antenna pairs were pointed either at the zenith or toward the NCP. In the former case, the LNAs that are normally attached to the feed antennas were disconnected and the VNA cables were attached directly to the feed outputs.  
When the antennas are pointed toward the NCP, the $S_{21}$ parameter is smaller 
{ due to blockage by the front dish antenna's reflector.}  
In this case, to increase the signal level, one of the LNAs remained attached to the output of one of the feed antennas in the antenna pair under test.

{ During the calibration of the VNA, we terminated the VNA ports with $50\,\Omega$ terminations and measured the noise floor to be about $-107\,$ dB.}

\section{Results}\label{result}

\subsection{Dependence on Baseline Length}
\begin{figure}[h]
    \centering
    \includegraphics[width=0.9\textwidth]{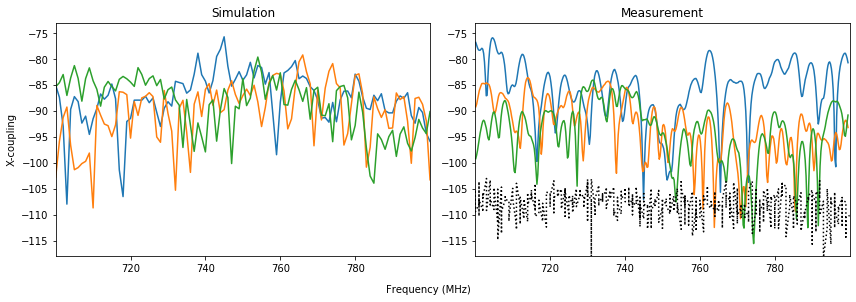}
    \caption{\textbf{Left:} { CST simulations of the} cross-coupling of baselines 2V-10V (blue), 2V-15V (orange), and 2V-8V (green) when the full array is pointing at the NCP. The baseline lengths are $\sim$\,11.5\,m, $\sim$\,20.3\,m, and $\sim$\,30.5\,m, respectively. { The simulations use the average of the $S_{21}$ and $S_{12}$ CST simulations.} \textbf{Right:} { VNA} measurement of cross-coupling ($S_{21}$) of 2V-10V (blue), 2V-15V (orange), and 2V-8V (green) { with dishes pointed toward the NCP. The black dotted line shows the noise floor for the VNA measurements. The $S_{21}$ measurements are described in Sec. \ref{sec:cross-coupling}. Time averaging was performed only on the noise floor measurements. Note that, although the simulations and VNA measurements do not match in detail, they are at about the same level. Note also that the frequency sampling of the two plots is different;  the CST simulation is sampled every 1 MHz, while the VNA measurement is sampled every 0.233 MHz. Ripples with spacing of a few MHz are visible in the VNA measurements.  The spacings are inversely proportional to baseline length and are consistent with the light travel times of the three different baselines.}  }
    \label{fig:full_array_baseline_length}
\end{figure}

\begin{figure}[h]
    \centering
    \includegraphics[width=0.9\textwidth]{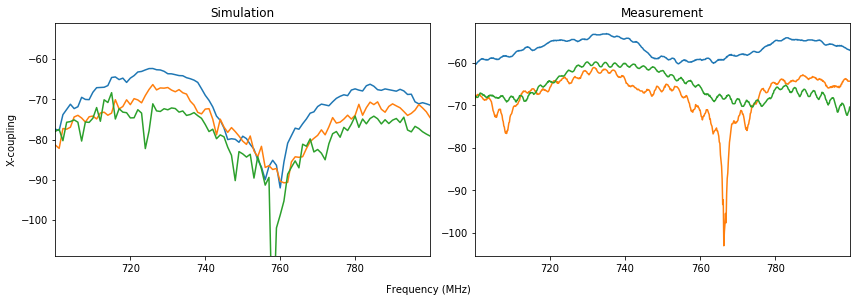}
    \caption{\textbf{Left:} Simulated cross-coupling of a zenith-pointing pair at 10\,m (blue), 20\,m (orange), and 35\,m (green) separations. \textbf{Right:} Measured cross-coupling of 2V-10V (blue), 2V-15V (orange), and 2V-8V (green) pointing at zenith. { In both cases the separations increase monotonically, but the distances are not exactly the same}. They should not be compared directly.}
    \label{fig:zenith_baseline_length}
\end{figure}

We first investigate how the cross-coupling depends on the baseline length. We expect a clear distance dependence of cross-coupling in sufficiently simple cases because the power radiated by an antenna scales inversely as the square of the distance from the antenna. { Both the simulations and the VNA measurements} with the full array pointing at the NCP { show cross-coupling at about the same level (Fig. \ref{fig:full_array_baseline_length}), although their detailed structure does not match.}. { However, in neither case} does the cross-coupling have a clear dependence on the baseline length. 
We think that this is due to the rotation of the dishes toward the NCP and the effects of blockage and reflection by dishes in the array and that we can be reasonably confident in the simulation results. In fact,  { neither the nightly mean visibilities (described below)} nor the cross-coupling measurements with the VNA show a noticeable dependence on separation. Moreover, { with the dishes pointed at the zenith} the simulated cross-coupling for a pair of dishes as well as the VNA measurements of pairs of dishes in the TDPA show much clearer baseline length dependence of the cross-coupling, as expected. (Fig. \ref{fig:zenith_baseline_length}).
It is noteworthy that the measured coupling between 2V-8V, which has longer baseline compared to 2V-15V, is at many frequencies higher than that of 2V-15V. This shows that accurate measurements of cross-coupling at this level is also quite difficult.

\subsection{Comparison of Simulated and Measured Cross-coupling}
\begin{figure}[h]
    \centering
    \includegraphics[width=0.95\textwidth]{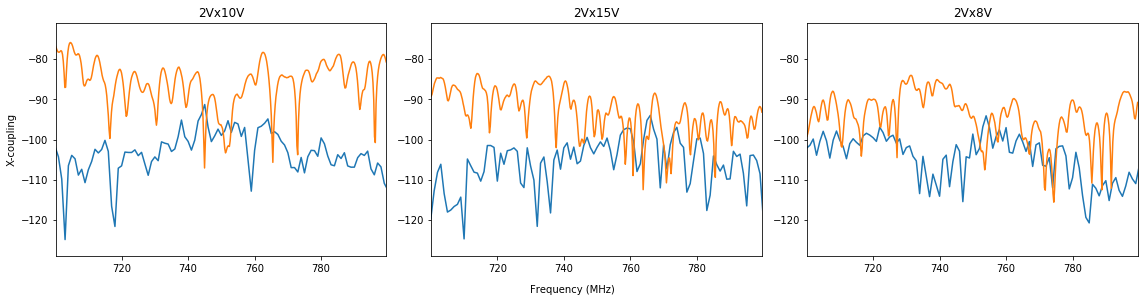}
    \caption{Comparison of the CST simulation of $S_{21}$ (blue) and the measured (with the VNA) $S_{21}$ (orange) for the corresponding baselines, with the antennas pointing at the NCP.} 
    \label{fig:SimVSMes}
\end{figure}

It is also interesting to compare directly the simulated and measured cross-coupling. In making Fig. \ref{fig:SimVSMes}, { we take the average} of the simulated $S_{21}$ and $S_{12}$, believing that they must be identical in theory due to symmetry but are not in our simulations. 
This is in line with our earlier claim that while the CST simulations accurately compute the level of coupling, the detailed frequency structure cannot be taken with certainty. We note that there may be systematic effects that may not have been accounted for in comparing the measurements and simulations, although the differences between them at some frequencies seem too large to be explained solely by such effects.

\subsection{Comparison to the Nightly Mean}
\begin{figure}[h]
    \centering
    \includegraphics[width=0.95\textwidth]{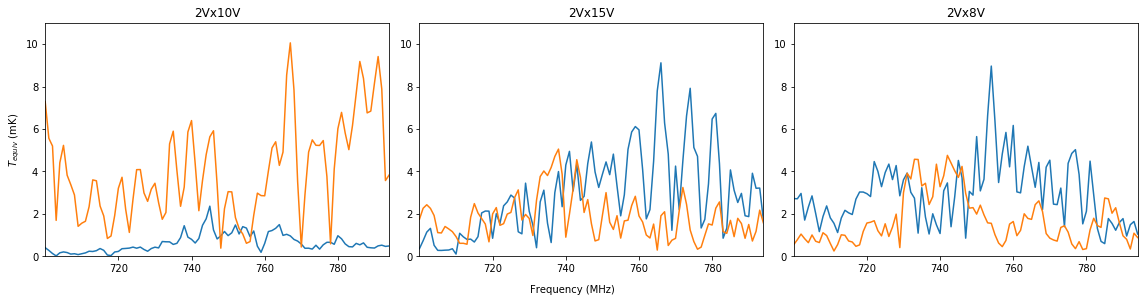}
    \caption{Comparison of visibilities in temperature units calculated from simulated (blue) and measured (with the VNA, orange) values of $S_{21}$. }
    \label{fig:SimVSMesTequiv}
\end{figure}

We also compare the simulated cross-coupling to the nightly mean visibilities from the 210 hour observation from 01/03/2018 to 01/11/2018 \cite{Wu2021}. 
{ A nightly mean visibility is the temporal mean of the calibrated visibilities in a specific interval of sidereal time.  This is recorded separately for each night and for each frequency channel.  The specific interval is chosen to extend over the sundown period as much as possible during an observation "run" (details in~\cite{Wu2021}).  Since this is a mean of visibilities calibrated with a bright source with known flux density all multiplicative linear responses of the telescope which are constant in time should be "calibrated out". This includes cable reflections. While the nightly mean visibilities are complex numbers only their absolute values are plotted.}  Eq. \ref{Eq:Visibility_temp_units_1} and Eq. \ref{Eq:Visibility_temp_units_2} are used to convert the cross-coupling to temperature units. The preceding discussion about the reliability of CST simulations and the measurements is inconclusive. Therefore, we compute the visibilities in temperature units from both the simulations and the measurements, as shown in Fig. \ref{fig:SimVSMesTequiv}, to be conservative in estimating the contribution of the crosstalk of receiver noise to the visibilities.

\begin{figure}[h]
    \centering
    \includegraphics[width=0.95\textwidth]{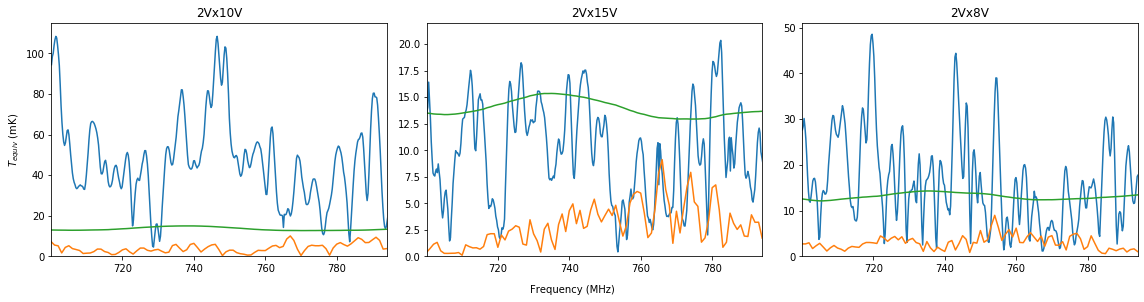}
    \caption{Comparison of the observed nightly mean { visibilities from the correlator} (blue),  cross-coupling (orange) { (based on the CST simulations of the $S_{21}$ parameter)}, and simulated sky contribution to the visibilities (green).}
    \label{fig:combined_signals}
\end{figure}

In comparing the contribution of the crosstalk to the visibilities to the nightly mean, we choose between the visibilities computed from simulated and measured cross-coupling the curves that have higher equivalent temperature. { In addition,} the contributions of the average sky to the visibilities are simulated and superposed. 
{ The simulated visibilities are computed using the {\bf JSkyMap}\footnote{JSkyMap git repository: \href{https://gitlab.in2p3.fr/SCosmoTools/JSkyMap}
{https://gitlab.in2p3.fr/SCosmoTools/JSkyMap}} 
simulation and map reconstruction software package, described briefly in 
\cite{PAON4_Zhang_2016}. JSkyMap can use a combination of point sources and a sky map corresponding to diffuse 
emission as inputs to compute visibilities; individual antenna or dish beams
can be specified as analytical shapes (Gaussian, Bessel, \ldots) or tabulated, direction dependent values.
The simulated visibility time streams for the NCP observations were computed for 
different baselines, using a frequency dependent Bessel shape beam, Haslam map extrapolated to the observation frequency, and NVSS sources, with flux extrapolated 
to the observation frequency, assuming a fixed value of the spectral index $\beta=-2$. NVSS sources with brightness $S_{21}>5\mathrm{Jy}$ and declination $\delta>15^\circ$, or $S_{21} >1 \,\mathrm{Jy}$ and $\delta>80^\circ$, or $S_{21} >0.5 \,\mathrm{Jy}$ and $\delta>85^\circ$ have been included. The sources within few degrees of the NCP dominates the simulated visibilities, with CasA contribution clearly visible, specially if CST simulated beam is used.} 
Fig. \ref{fig:combined_signals} shows that the cross-coupling is not the dominant contribution to the nightly mean visibility. In fact, the simulated sky has higher magnitude than the cross-coupling.

\section{Discussion}\label{discussion}
The challenge in accurately simulating and measuring the cross-coupling motivates antenna designs that suppress the cross-coupling as much as possible. Cross-coupling of -140\,dB combined with $S_{11}\lesssim -10$\,dB will give cross-coupling noise on the $0.01$\,mK level, making it similar to or lower than the expected HI signal. However, designing such an instrument may be very challenging.
\begin{figure}[h]
    \centering
    \includegraphics[width=0.5\textwidth]{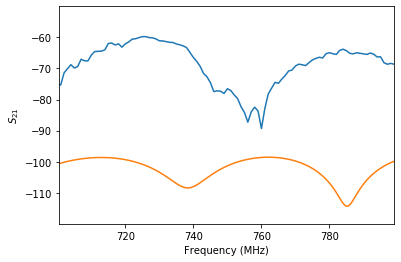}
    \caption{Simulated cross-coupling for the TDPA feed/dish design (blue) and an example of a feed/dish design optimized for low cross-coupling (orange), in both cases with a pair of antennas pointing at zenith.}
    \label{fig:TianlaiVSMoose}
\end{figure}
Here we present briefly the result from an attempt to mitigate the effects of crosstalk by design. With the feed and dishes described in \citet{JohnPFeed}, the simulation of a pair of zenith-pointing dishes at 10-m separation, which is the same setup as the simulation presented in Fig. \ref{fig:symmetry_pair_zenith}, gives between 10\,dB and 40\,dB lower coupling than the design adopted by the TDPA. One should note that the design in \citet{JohnPFeed} is optimized for an experiment with much wider bandwidth than the TDPA. By optimizing for a narrower bandwidth, we can reasonably expect lower cross-coupling. Otherwise stated, there is room for further improvement from the instrument design side. Still, the result from \citet{JohnPFeed} is roughly two orders of magnitude higher than desired. Consequently, the endeavor to better understand and model the cross-coupling must be continued.

\begin{figure}[h]
    \centering
    \includegraphics[width=0.9\textwidth]{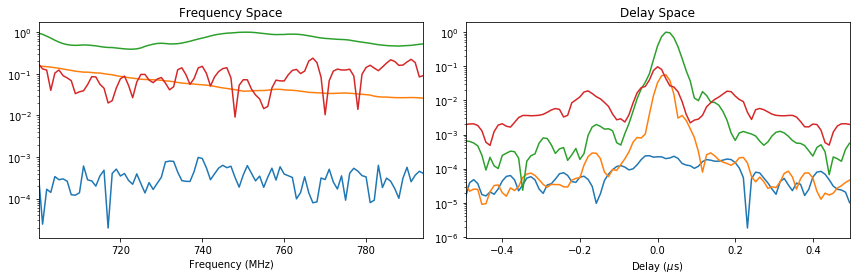}
    \caption{Visibilities from the expected HI signal (blue), polarized foreground (orange), unpolarized foreground (green), and crosstalk of receiver noise (red){, normalized to the strongest signal (unpolarized foreground),} for baseline 2V-10V in frequency space (left) and delay space (right). { The crosstalk of receiver noise is based on the measurement with the VNA.}}
    \label{fig:visibilities}
\end{figure}

To consider other possible strategies for cleaning the cross-coupled noise than direct simulation or measurement followed by subtraction, we compare the frequency structure and chromaticity of the visibilities contribution of the cross-coupling to the visibilities of the expected HI signal and foregrounds. The visibilities of HI and foregrounds shown in Fig. \ref{fig:visibilities} were computed using simulated maps produced by the CORA software package \cite{Shaw2014}. These maps were generated from 700\,MHz to 800\,MHz in 1\,MHz steps. For these visibilities, only simulations of Galactic synchrotron radiation and the HI signal were used. Beams at these same frequencies were constructed using E-patterns generated in CST simulations of the full TDPA. In particular, the E-pattern of dish 8 (vertical polarization) was rotated ro point toward the NCP. The rotated E-patterns of dish 8 were then used to form polarized (I,Q and U) power patterns. The visibility applied  corresponds to baseline 2V-10V.
The analysis for 2V-15V and 2V-8V gives similar figures as 2V-10V. The prescription for this procedure is described in \citet{Shaw2014}. 

In the visibilities shown, it appears that the crosstalk is quite chromatic, with its level dropping very little at higher delays. This means it would not be removable by fitting the data to smooth functions such as polynomials or Discrete Prolate Spheroidal Sequences (DPSS) as in \citet{Liu2020}. An option one can use to remove the cross-coupling noise with minimal signal loss is to perform  { singular value decompositions} (SVDs) on the visibilities, as studied by \citet{Kern2019, Kern2020}. One then removes the coupling effect by removing modes which vary slowly with Local Sidereal Time (LST). However, such a method may struggle on baselines oriented close to the North-South axis. One would expect both the crosstalk systematic and the sky signal to vary slowly with LST for such a baseline, making such a separation more difficult.

\section{Conclusions}\label{conclusion}
Our work shows that the cross-coupling of the receiver noise between the antennas in the TDPA is not the dominant source of the nightly mean visibilities in most cases. The relatively low level of cross-coupling suggests that it does not need immediate attention at the current stage. However, it is several orders of magnitude greater than the expected signal, making it a problem that the community must be mindful of when planning for future instruments. For the dominant source of the nightly mean in the TDPA visibilities, we point to the noise due to ground pickup as the next candidate for investigation. While sky signal has a meaningful contribution, it has a similar level across different baselines and hence is insufficient to explain the seemingly strong baseline dependence of the nightly mean.

We also conclude that CST simulations of cross-coupling using the integral equation solver with our setup tell us the level of cross-coupling with high confidence, but the detailed frequency structure should not be taken as accurate. { Nevertheless, we believe the cross-coupling is highly chromatic.  The dominant features in the delay spectra of the simulated $S_{21}$ for the different baselines are peaks at the expected delays.  Similarly, the frequency spectra of the VNA measurements of $S_{21}$ are dominated by ripples corresponding to light travel times of the different baseline lengths.}  Due to this relatively high chromaticity of the crosstalk, it would also be difficult to remove it with techniques similar to those developed for removing astronomical foregrounds. Consequently, the removal strategy for crosstalk that is generally applicable is to simulate or measure the crosstalk as accurately as possible and subtract its contribution to the visibility. Even though accurately simulating and measuring cross-coupling is difficult, the work to better understand and model such a noise must be continued. Design efforts to suppress cross-coupling as much as possible is also crucial for the success of 21\,cm intensity mapping experiments.

\section*{Acknowledgments}

We thank an anonymous referee, whose insightful comments have improved the manuscript considerably. We also are indebted to Calvin Osinga for his past work concerning CST simulations as a part of the group at UW-Madison and Danny Jacobs and Miguel Morales, for constructive discussions. The UW-Madison Center for High Throughput Computing provided  support for the lengthy CST computations. The Tianlai Pathfinders are operated with the support of the NAOC Astronomical Technology Center, and the Hebei Key Laboratory of Radio Astronomy Technology. Work at UW-Madison and Fermilab was partially supported by NSF Award AST-1616554. Fermilab is operated by Fermi Research Alliance, LLC, under Contract No. DE-AC02-07CH11359 with the US Department of Energy. Work at UW-Madison was further supported by the Graduate School, the Thomas G. Rosenmeyer Cosmology Fund, and by a student award from the Wisconsin Space Grant. Work at NAOC is supported by MOST grants 2022SKA0110100, NSFC grants 11473044, 11653003, 11773031, { 12203061} and 12273070, and CAS grant ZDKYYQ20200008. 
Authors affiliated with French institutions acknowledge partial support from CNRS (IN2P3 \& INSU), Observatoire de Paris and from Irfu/CEA. 
This document was prepared by the Tianlai Collaboration and includes personnel and uses resources of the Fermi National Accelerator Laboratory (Fermilab), a U.S. Department of Energy, Office of Science, HEP User Facility. Fermilab is managed by Fermi Research Alliance, LLC (FRA), acting under Contract No. DE-AC02-07CH11359.

\section*{Appendices}

\appendix{Solvers in CST}\label{app:cst}


For the simulations presented in this work, we experiment with two different solvers, Time Domain Solver (TDS) and Integral Equation Solver (IES). The experiment of TDS is motivated by \cite{Fagnoni2021} as it is used for the analysis of reflections from the HERA array, while our group has typically used IES for simulations with large geometry. TDS is based on the Finite Integration Technique. It applies numerical methods like the Perfect Boundary Approximation and the Thin Sheet Technique. These techniques allow for robust meshing in return for efficient memory usage. In comparison, IES is based on the Multilevel Fast Multipole Method (MLFMM). It uses surface meshing to analyze the frequency domain. The results of this solver contain information about the coupling between pairs of surface mesh elements. This process requires a lot of time and memory, but the  MLFMM is advantageous for problems involving large structures. Due to the difference in approaches TDS and IES take to solve the same problem, TDS requires all the ports in the simulation to be excited to yield a correct result whereas IES does not. The difference in method also implies that the accuracy settings for TDS and IES have different meanings. In both cases, the accuracy of the simulation is represented as ten raised to a negative power. For TDS simulations, this means that the simulation stops when the energy remaining in the system is the energy emitted by the signal times the accuracy setting, whereas in IES, the simulation stops when the percentage difference between successive iterations is smaller than the accuracy setting.

This leads to various reasons why IES is better for our purposes than TDS. TDS has advantages in systems with translational symmetry, which is true for the configuration of HERA that is presented in \cite{Fagnoni2021}. However, the TDPA is radially symmetric; moreover, this symmetry is broken by pointing the dishes toward the NCP. IES handles such rotations of the array elements, which we need to account for in TDPA because the observation we compared to is performed with dishes pointing at the NCP.  HERA did not use this method because HERA is zenith-pointing.
In addition, since the results of the TDS simulation relies on one calculation that takes weeks to complete in our computation whereas fragmentation by frequency samples is possible for IES simulations, simulations with IES are much less likely to be interrupted by computing and other practical factors. This difference is particularly dramatic if many simulations can be done at the same time, as in our case with multiple licenses and the aid of high throughput computing clusters.

\bibliographystyle{ws-jai}
\bibliography{refs.bib}



\end{document}